# How network temporal dynamics shape a mutualistic system with invasive species?

*Andrzej Jarynowski[1], Fco. Alejandro López-Núñez[2], Han Fan[3]*

1 Smoluchowski Institute, Jagiellonian University., Cracow, Pl.   andrzej.jarynowski@sociology.su.se
2 Centre for Functional Ecology, University of Coimbra, Pt.    lnfran85@gmail.com
3 Chalmers University of Technology, Gothenburg, Se.   han.fan@foxmail.com

*Abstract*

Ecological networks allow us to study the structure and function of ecosystems and gain insights on species resilience/stability. The study of this ecological networks is usually a snapshot focused in a limited specific range of space and time, prevent us to perceive the real dynamics of ecological processes. By definition, an alien species has some ecological strategies and traits that permit it to compete better than the native species (e.g. absence of predators, different bloom period, high grow rate, etc.). Plant-pollinator networks provide valuable services to whole ecosystems and the introduction of an alien species may have different effects on the native network (competitive facilitation, native species extinction, etc.). While scientists acknowledge the significance of network connectivity in driving ecosystem services, the inclusion of temporary networks in ecological models is still in its infancy. We propose to use existing data on seasonality to develop a simulation platform that show inference between temporality of networks and invasions traits. Our focus is only to pick up some simple model to show, that theoretically temporal aspect play a role (different extinction patterns) to encourage ecologist to get involved in temporal networks. Moreover, the derived simulations could be further extended and adjust to other ecological questions.

## 1. Introduction

Network theory is useful when it comes to study nature from a systems perspective, and there are several examples in which it has helped understanding the behavior of complex systems. Ecosystems are example in which a network perspective is important to understand systems behavior (Fig. 1). Such networks are defined by nodes-species or individuals and links-interactions like as seed dispersal or intraspecific competition. Many studies have been done on biodiversity, invasion processes and mutualistic interactions with network approach (Ferrero *et al.*, 2013; Heleno *et al.*, 2013; Olesen, Eskildsen & Venkatasamy, 2002). It is known that invasive species represent one of the principal threats regarding the functioning and stability of ecosystems (Chornesky& Randall, 2003) and they can modify ecological interactions (Traveset& Richardson, 2006). However, agents (species or individuals) in those systems also interact with each other at different time-scale and in different ways (Fig. 2). Temporal patterns exist in many ecological systems such as predator-prey's food webs, host-parasite's webs, etc. Some works with pollination networks have demonstrated that important ecological factors have a strong seasonal variability (Basilio *et al.*, 2006, Olesen *et al.*, 2008), but in general, thedynamics from this perspective have not been studied in detail and an integrative framework is missing. Temporal data is available, but the study of the effects on species relationships (e.g., coexistence, persistence, etc.) is underexplored and need some larger attention. In this work we propose a dynamical model that integrates competitive interactions and mutualistic species, combining network theory with rule-base modeling. We will apply and test our model with artificial island's plant pollinator community to understand process. We propose that our model represents a general and scalable approach to study a wide range of mutualistic network interactions (like seed dispersal also) in order to understand their structural and dynamics properties allowing to make a "predictions" (e.g., how system react to a introduction of alien species).

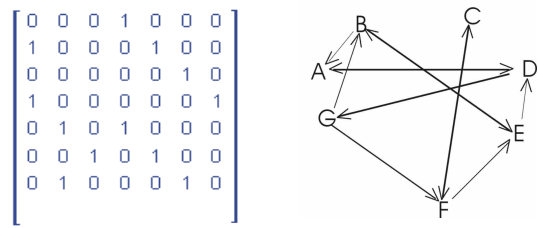

**Fig. 1.** Visualtization of network based on theoretical links - interactions between assets - species.

There are few methods to represent reduced connection matrix (network). For instance, only statistically significant (with given level) links remain. On the other hand, it is possible to provide treshold that makes strong links survive in the network. Plant-pollinator mutualistic networks have many proprties. They are asymmetric in their interactions: specialist plants are pollinated by generalist animals, while generalist plants are pollinated by a broad range involving specialists and generalists (Abramson *et al.*, 2011). This structure of seasonal interaction is also difficult to understand, so it is reasonable to introduce temporal networks instead of qualitative description (Fig. 2).

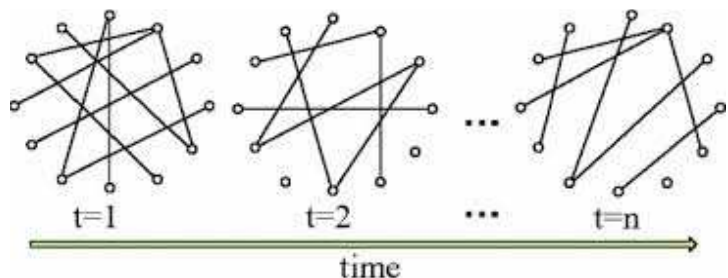

**Fig. 2.** Representation of temporal network.

We use temporal networks approach within Vensim system dynamic difference equations environment (Jarynowski & Serafimovic, 2014). Vensim is simulation software for improving the performance of real systems. System dynamics is the study and analysis of dynamic feedback systems using computer simulation. We provide simulation of the implications of all the relationships that have been specified for the variables in a plant-polinator model. Simulations in Vensim result in the behavior (over time) of all the variables in the model to test: how synchronization lead to stabilization of the network? Or are extinction of the species possible and if their create cascades?

## 2. Methods

To design our study model (see Fig. 3), we revisited the Bastolla *et al.* 2009 population dynamic set of differential equations.
- We use Euler approx. to solve proposed difference equations in Vensim.
- We choose the set of parameters to make system exist.

**Tab. 1.** Populations and parameters in our model

| Name | Function |
|---|---|
| F1 | Flower (Plant) – native and specialist with short flowering period |
| F2 | Flower (Plant) – native and generalist with long flowering period |
| F3 | Flower (Plant) – alien and generalist with the longest flowering period |
| I1 | Insect (Pollinator) – generalist with long breeding period |
| I2 | Insect (Pollinator) – specialist with long breeding period |
| I | Predator of Insect population |
| $af=af_1=af_2$ | Birth rate for native flowers |
| $ai$ | Birth rate for insects |
| $af_3$ | Birth rate for alien flower |
| $df$ | Death rate for flowers |
| $di$ | Death rate for insects |
| $gf$ | Mutualistic profit factor for flowers |
| $gi$ | Mutualistic profit factor for insects |
| $bf$ | Competition factor for flowers |
| $pred$ | Predation factor for insects |
| $c$ | Adventage in mutualistic interactio of being specialist plant |
| $T_{ki}$ | Synchronization in seasonality. If $k$ and $i$ are active $T_{ki}=1$, otherwise $T_{ki}=0$ |

Assumptions:
- Limited space (300 cell for plant settlement).
- Constant and equal natural death and reproduction rates for native plants.
- Invasive plant has bigger reproduction rate than native ones.
- Constant and equal natural death and reproduction rates for insects.
- Insect's population controlled by their predators.
- Constant and equal (symmetric) competition rates of plants.
- No direct competition between insects.

Our system is described by with labels in Tab. 1:

$$\frac{dF_i}{dt} = \underbrace{(af_i - df_i)F_i}_{\text{Natural birth/death}} - \underbrace{\sum_j bf * F_i F_j}_{\text{Competition}} + \underbrace{\sum_k gf * F_i I_k * T_{ik} * c_i}_{\text{Mutulistic profis}}$$

$$\frac{dI_i}{dt} = \underbrace{(ai - di)F_i}_{\text{Natural birth/death}} - \underbrace{pred * II_i}_{\text{Predation regulation}} + \underbrace{\sum_k gi * F_k I_i * T_{ik}}_{\text{Mutulistic profis}}$$

Scenarios:
- Static (no seasonality all inter and intra specific interaction constant with time).
- Dynamic (seasonality included in interspecific interaction, mutualistic advantage can appear only if both reproductabily active – birth rate).
- Changeable tradeoff rate for specialism.
- With or without invader plant.

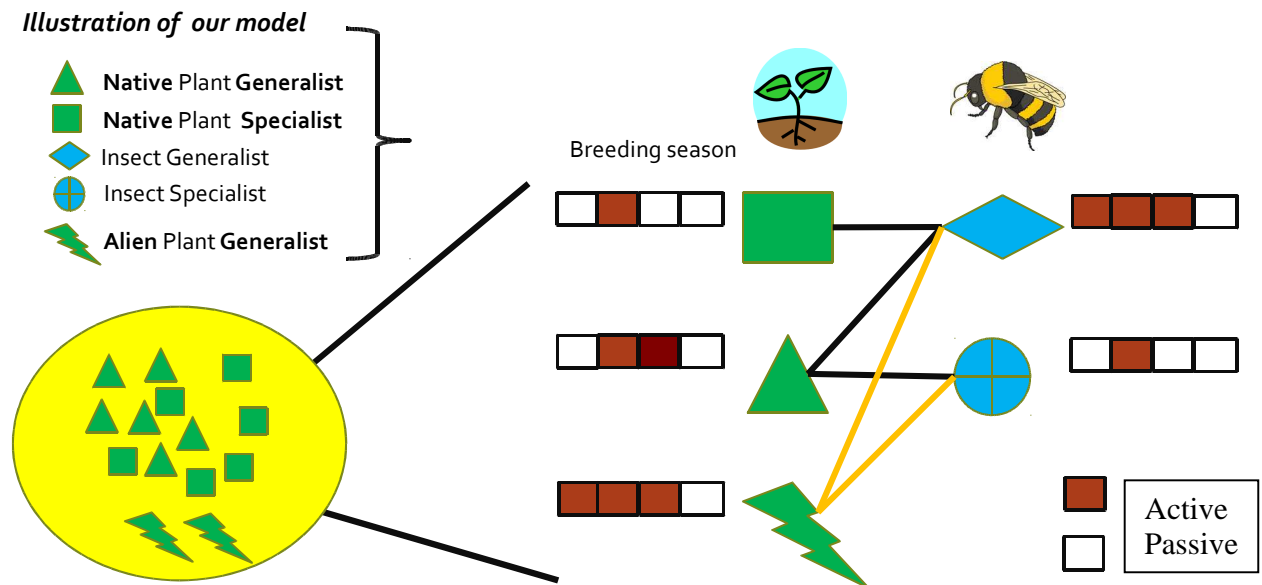

**Fig. 3.** Schema of our proposed plant-pollinator study model.

## 3. Preliminary results

In the static scenario, when the alien species has a double valor of birth-rate ($af = 2$) than the native species ($af = 1$), our results show that the introduction of a generalist alien plant causes the extinction of the populations of the specialist native plant species, and at the same time, the invasive plant goes extinct and the generalist native plant species and the different pollinators are unaffected (Fig. 4). But if the alien species' birth-rate is four times high ($af = 4.375$), all native plant species goes extinct, and only survived the alien species and the insects are unaffected (Fig. 5). By other hand, if the plant competition factor is slightly greater ($c = 1.537$) and also the alien species' birth-rate is slightly less ($af = 3.125$), the native plant species and the specialist insect goes extinct, but the alien plant survives (Fig. 6). When we introduce the seasonality in the model, the results show that only survive the generalist native plant and their specialist pollinator. The other species collapse (Fig. 7). Due to limited space assumption, initial condition of abundance of invasive plant plays a role. If initial invasive plant number is too few even of advantage in reproduction rate, new specie could go extinct, due to "majority domination". A initial population ratio between native and alien play a role. However the fitness adventage of the alien plant depend of its birth-rate, and therefore, this birth-rate is the cause of the extintion of the native plants. If we maintain a few alien population size and increment its birth rate, the effect can be seen.

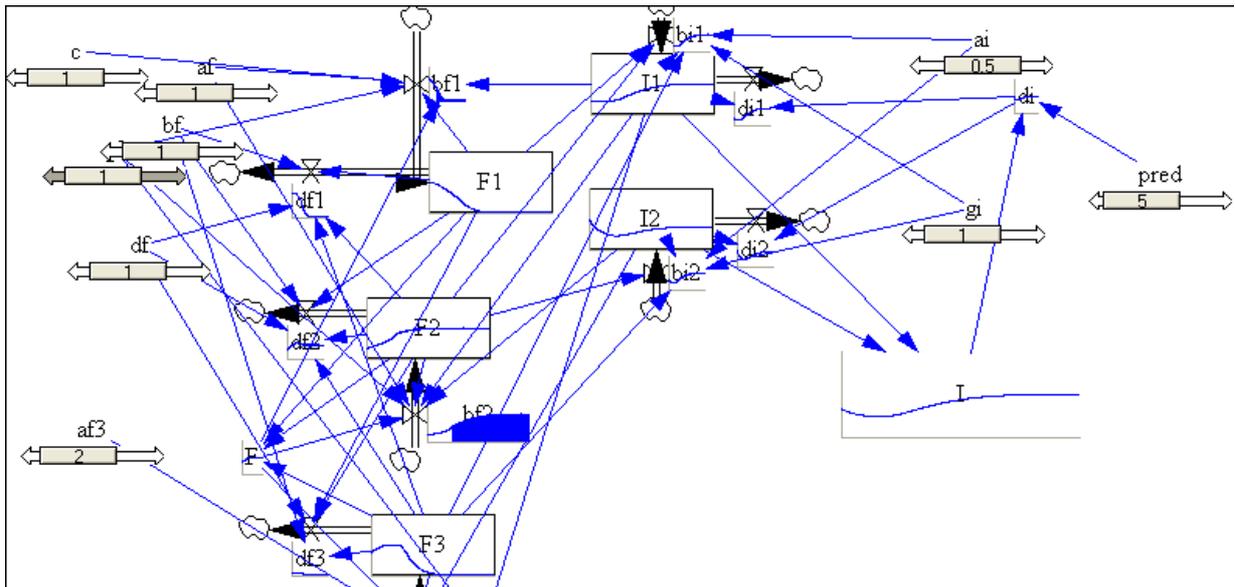

**Fig. 4.** Preliminary results – static case trajectories. Scenario when specialist plant goes extinct as well as invasive one.

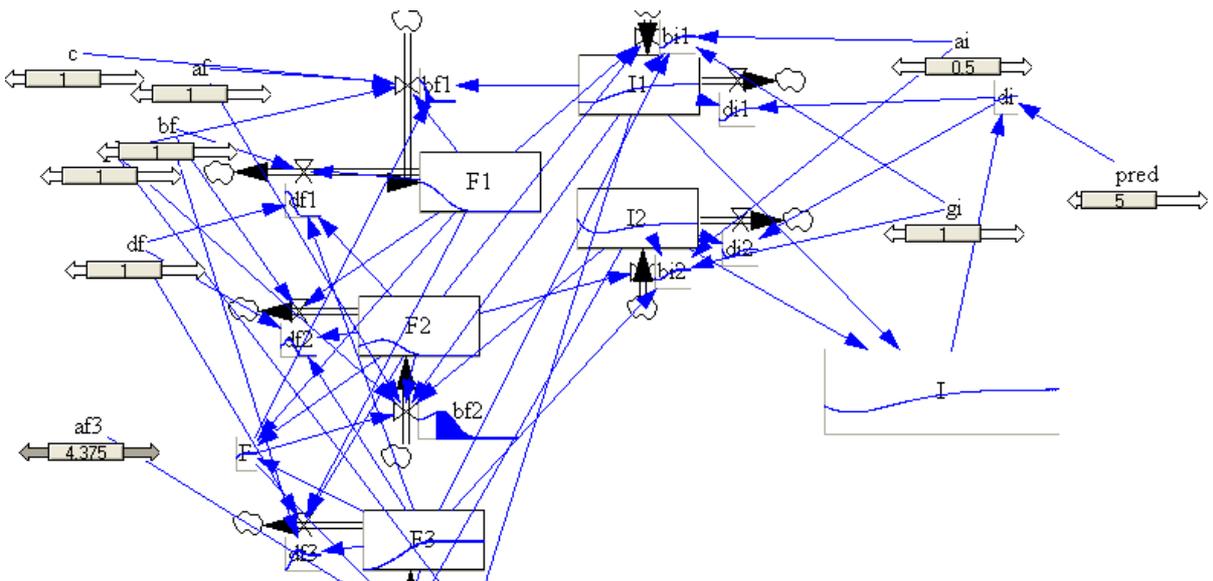

**Fig. 5.** Preliminary results – static case trajectories. Natives go extinct and invasive plant takes all space.

Technically, Vensim is solving systems of lumped ordinary difference equations. To undersand pictures of the stock and flow diagram in time (Fig 4-7) let introduce some concepts of modeling in Vensim (Popova *et al.*, 2011). All such processes can be characterized in terms of variables of two types, stocks (population size F1, F2, F3, I1, I2) , big arrows (flows of birth and deaths), small arrows (rates of change due to procces and interactions). With the discrete (static or dynamic) and deterministic flow assumptions, a ecosystem isbasically modeled as a „plumbing" system. We present graphical sensitivity view and the results of the simulation are graphically overlaid on the model. Scales with bar sliders are used to represent changes that can be made to constants (like valor of birth-rate *af* or the plant competition factor *c*). Graphs are used to represent the output or impact on model variables (time evolution of population size of F1, F2, F3, I1, I2). In dynamical case (Fig. 7) seasonality patterns (Fig. 2) are applied to connect or disconnect networks for changing seasons of the years.

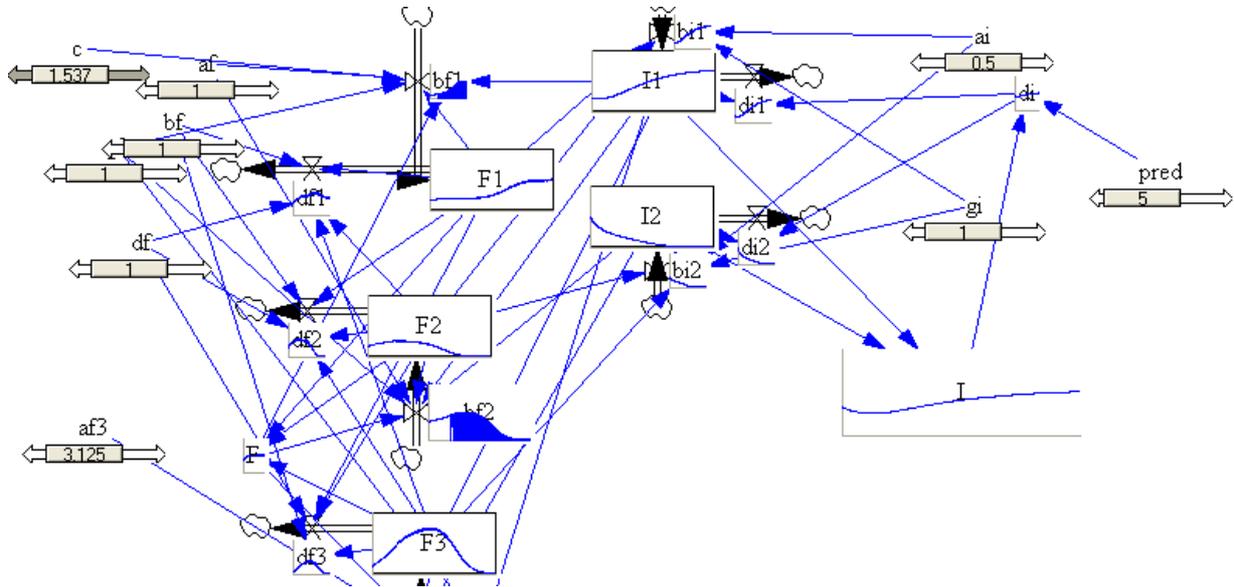

**Fig. 6.** Preliminary results – static case trajectories. Native generalist and invasive plantgo extinct and specialist plant takes all space. Specialist insect goes extinct indirectly.

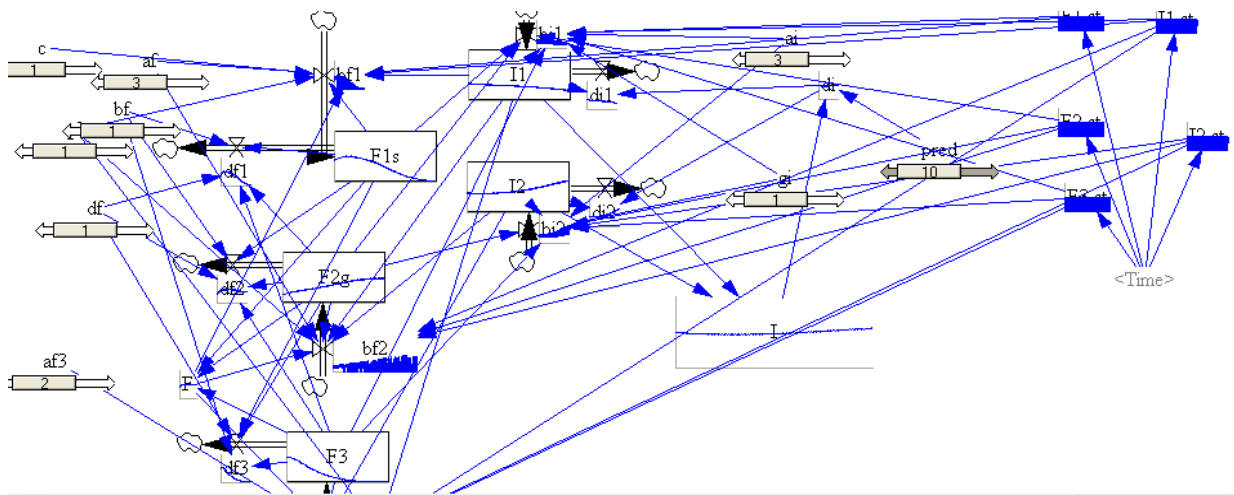

**Fig. 7.** Preliminary results – dynamic case trajectories. Scenario when specialist plant goes extinct as well as invasive one, generalist insect is also going to extinct.

## 4. Research plan and future works

Our objectives are to introduce of temporal dynamics to ecological system by studying plant-pollinator network on an island with limited space. We ask for such a question - How temporal aspect (seasonality) of the networks with native and invasive species affect ecosystem in terms of:

- Stabilization of the network;
- Extinction of the species.

Some additional research should be done:
- To test different seasonal traits (combination of permutation of time vector) to find if extinction or coexistence patterns depend on them.
- To research the majority effect of native vs invasive species.
- Try empirical datasets (e.g. Azores and Mauritius Islands and other small islands with small networks max. 10 by 10).

## 5. Conclusions

Identifying and quantifying the factors that are important to accurately simulate population's dynamics throughout a dynamic network with of without invasion is an ongoing challenge.In this study we show the importance of the seasonality in different species population dynamics. By other hand, using different species traits we can simulate and predict the behavior of each node in the network over a time range. We believe that our studies can address the importance of temporal patterns for the ecological scientific knowledge production, while many theoretical researches were done in physics and mathematics. In future, we would like to run a multi-pronged approach using individual-based computer simulations, and cellular automata to introduce geographical initiation of new plants.

## Acknowledgments

This report is an result of discussion during the spring 2014 on workshop organized by Icelab, Umea University: Networks in Ecology about combining network science, dynamical systems and ecology. Many thanks to Serguei Saavedra, Jan Haerter and Magnus Lindh.